\documentclass{emulateapj}

\usepackage{url}
\usepackage{graphicx}
\usepackage[colorlinks,urlcolor=blue,citecolor=blue,linkcolor=blue]{hyperref}

\begin{document}

\title{A Systematic Review of Strong Gravitational Lens Modeling Software}
\author{Alan T. Lefor}
\author{Toshifumi Futamase}
\author{Mohammad Akhlaghi}
\email{alefor@astr.tohoku.ac.jp}
\affil{Astronomical Institute, Tohoku University, 6-3 Aramaki, Aoba-ku, Sendai 980-8578, Japan}

\begin{abstract}
Despite expanding research activity in gravitational lens modeling, there is no particular software which is considered a standard. Much of the gravitational lens modeling software is written by individual investigators  for their own use. Some gravitational lens modeling software is freely available for download but is widely variable with regard to ease of use and quality of documentation. This review of 13 software packages was undertaken to provide a single source of information. Gravitational lens models are classified as parametric models or non-parametric models, and can be further divided into research and educational software. Software used in research includes the GRAVLENS package (with both gravlens and lensmodel), Lenstool, LensPerfect, glafic,  PixeLens, SimpLens, Lensview, and GRALE. In this review, GravLensHD, G-Lens, Gravitational Lensing, lens and MOWGLI are categorized as educational programs that are useful for demonstrating various aspects of lensing. Each of the 13 software packages is reviewed with regard to software features (installation, documentation, files provided, etc.) and lensing features (type of model, input data, output data, etc.) as well as a brief review of studies where they have been used.  Recent studies have demonstrated the utility of strong gravitational lensing data for mass mapping, and suggest increased use of these techniques in the future. Coupled with the advent of greatly improved imaging, new approaches to modeling of strong gravitational lens systems are needed. This is the first systematic review of strong gravitational lens modeling software, providing investigators with a starting point for future software development to further advance gravitational lens modeling research.

\end{abstract}

\maketitle

\section{Introduction}
Gravitational lensing has great promise to provide new insights into the structure and history of the universe. Gravitational lensing has yielded many exciting results by mapping dark matter distributions, and the recent use of strong gravitational lensing data has added a new dimension to this research \citep{Coe2010}. Gravitational lensing is a very active area of investigation, and research is highly dependent on computer modeling. In some areas of contemporary astrophysics research, there is software that is a \emph{de facto} standard for many investigators (e.g. SExtractor \citep{Sextractor} \footnote{\url{http://astroa.physics.metu.edu.tr/MANUALS/sextractor/}}, GALFIT \citep{Galfit} \footnote{\url{http://users.obs.carnegiescience.edu/peng/work/galfit/galfit.html}}, etc.). While gravitational lens modeling software has been written, there are no standards  and no easily accessible source of information about existing software. The lack of standard software may be a virtue of the gravitational lensing community, allowing more flexibility and creativity. The lack of a single standard program makes it more important to compare existing software used for modeling strong gravitational lenses. Information regarding existing software will be helpful to those developing new approaches and interfaces. 

This review was undertaken to identify available strong lens modeling software, and review the installation, use, and the nature of inputs and data outputs. This paper  serves as a guide to available software and  provides useful information to both new and established investigators in this field. The availability of  source code may be a useful starting point for anyone writing their own modeling software. 

This paper is organized as follows. In section \S \ref{Classif}, we review the classification of gravitational lens models and the methodology used to review the available software. In section \S \ref{Res}, we review eight software packages that have been used extensively in gravitational lens research. Following this, in section \S \ref{Educ}, we review five programs that are useful in education for gravitational lensing.  In section \S \ref{Compare} we discuss several factors of importance in selecting and comparing available software and in section \S \ref{Concl} we make suggestions for the next generation of software to support  future  gravitational lens research based on this review.

\section{Classification and Review Methodology} \label{Classif}

Modeling of gravitational lenses starts with a list of observables such as relative positions of the components, relative fluxes of the images, time delays between the images and other lens properties. Gravitational lens modeling can be considered a ``forward'' or a ``reverse'' problem. In the forward problem, images are predicted based on a known source and lensing mass. More commonly, the reverse problem is considered, using observed images to reconstruct a model of the mass density based on the images, usually approached using non-parametric methods.  

In parametric models, a clear physical parameterization is used to construct the model from the outset, while non-parametric models are often  ``grid-based'', although there are other methods used. Both parametric and non-parametric models are valuable, since some features of individual lenses are model independent while others are very model dependent \citep{SimplensSaha2003}. A major distinction is whether the calculation is ``model-based'' (parametric) or ``model-free'' (non-parametric) at the start of the process  \citep{Coe2008}. A more recent approach to model classification avoids the confusion of parametric vs. non-parametric (since all models use parameters) and classifies models as light traces mass (LTM) or Non-LTM \citep{Coe2010}.

\subsection{Parametric Models}
Parametric simulation models are generally used to solve the ``forward'' problem, taking a source and lensing mass, and then predicting the resulting image. Parametric models are also known as simply-parameterized models, and assume a physical model which fits the data with relatively few defined parameters \citep{Julio:2007}. In parametric models, the data is fitted to a physical object (e.g. Point Mass, Singular Isothermal Sphere, Singular Isothermal ellipsoid, De Vaucouleurs model, Navarro-Frank-White Model, etc.) and a model of the lensing mass made using that physical object to predict the effect on the light from the source.

\subsection{Non-Parametric Models}
Non-parametric models are often used to solve the more complex ``reverse'' problem, also called lens inversion, of taking a lensed image and from it, predicting the nature of the lensing mass. The lens inversion problem is complicated by the fact that there are huge degeneracies in the parameter space which make several models able to fit observed data \citep{Meneghetti}. The degeneracy problem is inherent in lens inversion, because the constraints on the potential are local \citep{2009AAAllard}. The principle underlying non-parametric gravitational lens models is that the effective lens potential and the deflection equations are linear functions of the surface density \citep{KochanekSaas}. These models reconstruct the mass distribution or lens potential as a map defined on a grid of pixels \citep{Julio:2007}. By using a large number of parameters these models are very flexible, but conversely, the large number of parameters can lead to over-fitting the data. The parameters are usually in the form of basis functions, and given the large number used, the term ``non-parametric'' is somewhat paradoxical, although it refers to the lack of a discrete physical model being used in the solution of the problem \citep{LiesenborgsTh}. It is only recently that lens inversion is studied using strong gravitational lensing data.

Lens inversion methods are classified into 3 types: (i) model dependent reconstructions, (ii) potential reconstruction on a grid and (iii) expansion of the potential functions. Degeneracy must be dealt with using any of these techniques \citep{2009AAAllard}. Further examples of these techniques include the maximum entropy method \citep{LensMEMAlg}, genetic algorithms \citep{2005PASABrewerLewis, LiesenborgsTh}, Bayesian analysis \citep{MNR:MNR10733, 2006ApJBrewerLewis}, the semilinear technique \citep{2003ApJWarren}, and perturbative reconstruction \citep{2009AAAllard}. 

The ideal lens inversion algorithm (i) should be free of assumptions regarding mass or luminosity distributions, (ii) should not depend on prior information, (iii) should not produce models that are physically impossible, (iv) should be free of uncontrollable parameters, and (v) should be extendable to any kind of data \citep{Liesenborgs2006}. The fact that there are so many techniques being used to investigate lens inversion, indicates that there is no single ``best'' technique. The theoretical underpinnings of these various methods have been reviewed and shown to be essentially  different methods of doing the same thing \citep{2006ApJBrewerLewis}.

\subsection{Education and Research Modeling Software}
Gravitational lens modeling software can also be  arbitrarily classified into packages used for education and those used for research. Those classified for research in this review have been used in published studies of gravitational lensing. Those classified as educational have not been used for studies. 

\subsection{Review Methodology}
There are two main goals when evaluating gravitational lens modeling software \citep{Wayth2006}. The first is to determine if the software can recover the lens model parameters of a known lens, and with what accuracy. The second regards the accuracy of the source reconstruction. The accomplishment of these two goals must be considered against the background of software usability and efficiency. 

Gravitational lens modeling software was identified by using search engines on the Internet as well as searching the literature on arXiv.org. The software reviewed here was chosen based on availability for download and the ability to install and execute the programs. Another resource for gravitational lens modeling software is the Astro-Code Wiki, which includes links for several of the modeling codes reviewed here \citep{Astrocode}. Information regarding release dates and versions was obtained from the web sites. Each of the software packages described here was downloaded, compiled (when necessary) and installed. Sample data files were used when available, and output shown here is directly from the downloaded version. All of the software functioned as described by their developers. Documentation was also downloaded directly from the internet. This review was not intended to be all inclusive; other lens modeling packages are not reviewed.  Several of these programs, although web sites for download were identified on line or in publications, were no longer available for download. Others had significant issues when attempting to compile and execute the programs, precluding their use. There are likely other software packages being used by individuals but not available for download on the internet. The software reviewed here is representative of what is available in regard to algorithms, types of models available and feature sets. Features of the software reviewed relating to installation and use are summarized in Table 1. Features of the software relating to lensing models and algorithms are summarized in Table 2.

\begin{table}[t]

\centering

\begin{tabular}{c c c c c}
\hline
Package & Year & Source & Exec & Platform \\
\hline
gravlens / lensmodel & 2008 & No & Yes & PPC, Linux, OS X \\
Lensview & 2008 &Yes & No & OS X/Linux\\
Lenstool & 2006 & Yes & No & OS X/Linux \\
LensPerfect & 2007 & Yes & No & Python \\
glafic & 2012 & No & Yes & OS X/Linux \\
PixeLens & 2007 & Yes & Yes & Java \\
SimpLens & 2003 & Yes & Yes & Java \\
GRALE & 2008 & Yes & No & OS X/Linux \\
\hline
GravLensHD & 2011 & No & Yes & iOS \\ 
G-Lens & 1998 & No & Yes & DOS \\
Gravitational Lensing & 2002 & No & Yes & Win/HP-49 \\
lens & 2002 & Yes & No & MATLAB \\
MOWGLI & 2013 & No & Yes & Java \\

\end{tabular}
\label{table:Packages}
\caption{A summary of software features of gravitational lens modeling software. Source indicates that the source code is provided, Exec indicates that a downloadable executable file is provided and Platform indicates the computing platform that the executable works on. (PPC=PowerPC, Dates are the current version available, Exec=available executable code)}
\end{table}


\section{Research Software} \label{Res}

Strong gravitational lens modeling software for research use is  in this category. The eight software packages described here have been used to analyze experimental data in published studies.

\subsection{gravlens / lensmodel}
\subsubsection{Software features}

The GRAVLENS- Software for gravitational lensing package was first released in 2001, and is now in version 1.99 dated 2008\footnote{\url{http://redfive.rutgers.edu/~keeton/gravlens/}}. The software is available as a download, with executable files provided for PPC and Linux architectures. Source code is not available. Documentation is also available as a 101 page user manual \citep{GravLensManual}. The lensmodel program is available as a download, with executable files provided for PPC and Linux architectures. Source code is not available. The lensmodel software uses the  gravlens kernel and adds functionality, and was also described by \cite{GravLens}.  A tutorial is provided with the user manual. The user interface is character based and input is through a text file with text commands. There are two on-line tutorials that illustrate many of the features of the software. While gravlens includes basic lensing calculations, lensmodel includes added routines to model strong lenses \citep{GravLensManual}.

\subsubsection{Lensing features}

 The lensmodel software uses a parametric model, and has been used in a number of research studies  \citep{Cohn2000, Allam2007, Alba2007, Lin2009, Smith2005, Aazami2006}. GRAVLENS was first described along with a new algorithm that allowed calculation of lensing parameters for a generalized mass model, allowing implementation of multiple parameterized models \citep{GravLens}. GRAVLENS is a sophisticated software package that is accompanied by a large catalog of parametric lensing models \citep{GravlensCat}. Both programs in GRAVLENS includes a wide range of basic lensing calculations which are based on tiling of the image. The heart of the code is a general algorithm for solving the lens equation and is fully described in \cite{GravLens}. This algorithm involves tiling the image and source planes, and using these tiles to determine the number and approximate positions of all lensed images associated with a given source. Performing a calculation requires specifying the details of the tiling and the parameters of the mass model.  The code uses a polar grid centered on the main galaxy. Default values work with most calculations. Optionally, a critical curve grid can be used in which critical curves are used to determine where to place radial zones. Various options for plotting the output are available, and the output is a file of macros for input into the SuperMongo plotting program. The lensmodel software expands on the capabilities of gravlens, to include routines that make it easy to fit models to observed lens systems. The central portion of lensmodel computes a $\chi^2$ value for a set of models. A wide variety of lens models are available in the catalog \citep{GravlensCat}. The software can also calculate the Hubble constant using time delays. The available tutorials are very helpful for illustrating and explaining the complex commands that are used in gravlens and lensmodel. 

Gravlens was used to model strong gravitational lenses in an interesting study which examined the cusp and fold relations as a gauge of substructure \citep{Aazami2006}. This was a theoretical investigation to look at the sensitivity of these relations to the presence of substructure in the lens. The authors found that the fold relation is a more robust indicator of substructure than the cusp. Gravlens provided an excellent analytical tool for this study. A newly discovered elliptical galaxy at z=0.0345 was investigated and modeled with gravlens/lensmodel \citep{Smith2005}. The model was created with a SIE, and was found to be in close agreement with the light distribution observed. 

In a study of a bright strongly lensed z=2 galaxy in the SDSS DR5, Lin and colleagues used Lensview to model a lensing system \citep{Lin2009}. However, since Lensview uses the full image information, they employed gravlens/lensmodel to fit an SIE model using only the image positions. The resulting model showed a very good fit to the image positions. This study is one of the very few \emph{direct comparisons} of strong gravitational lens modeling software in the literature. 

The ten image radio lens, B1933+503 was modeled using gravlens by Cohn and coworkers \citep{Cohn2000}. The mass distribution of this system was modeled with a wide variety of parameterized ellipsoidal density distributions. The models were constrained using the relative positions of the lens galaxy and the lensed images and the flux ratios between the images. The mass distribution was concluded to have an approximately flat rotation curve based on this analysis. The gravlens program was used to model the cluster lens MS0451.6-0305 by Alba and colleagues \citep{Alba2007}. A simple elliptical lens model (SIE) with external shear was used. They used an NFW profile for the cluster mass distribution which was consistent with observations. The model reproduced the positions of the Extremely Red Object images and the radio images very well. This model was used to test if the configuration of the observed radio emission could be understood as a result of gravitational lensing, and was able to explain the morphology of the radio map as a result of the three lensed background sources. Gravlens was also used to model a strongly lensed Lyman Break galaxy at z=2.73, identified in the SDSS DR4 imaging data \citep{Allam2007}. The authors assumed a SIE and used gravlens/lensmodel to perform fits to the data. The fitted values showed excellent agreement with observed values from the SDSS DR4 database. 

\subsubsection{Summary}
Both gravlens and lensmodel require no installation as they are provided as executable binary files. They are somewhat complicated to use because of the character based interface and extensive command set. They would benefit  from a graphical user interface which is described in a web site but the code has not been available \citep{AlfaroWeb} \footnote{\url{http://cinespa.ucr.ac.cr/software/xfgl/index.html}}. The lack of source code is also a limiting factor for those wishing to study the computational techniques used. Finally, the programs are limited by requiring an outside plotting package to view graphical results. The required plotting package is expensive and somewhat outdated. Both gravlens and lensmodel have an extensive catalog of available mass models that can work with complex datasets, and perform very well in regard to comparing the models generated with observational results. Overall, they are excellent strong gravitational lens modeling software packages.

\subsection{Lenstool}

\subsubsection{Software features}
Lenstool was first written in 1993, and is now released as version 6.7.1, dated 2006\footnote{\url{http://www.oamp.fr/cosmology/lenstool/}}. Lenstool is distributed as source code with dependencies on WCSTOOLS, PGPLOT and CFITSIO, and the latest version is available on the Lenstool Project web site \citep{LenstoolWeb}.  Installation is somewhat tedious but can be accomplished with available standard libraries, and can be installed in Linux or OS X. The user interface is character based and input is through a text file with text commands. A 61-page user manual as well as a 41 page document entitled ``Lenstool for Dummies''  are available for download \citep{LTDummies}. The documentation is excellent for this complex and comprehensive strong gravitational lens modeling software.

\subsubsection{Lensing features}
Input to Lenstool is a character based input file. Each line consists of a command and appropriate data. Keywords are either first identifiers or second identifiers. A number of input files are necessary. The first is a PAR file, containing the basic parameters for the model, a list of the arcs for which Lenstool will predict counter images, and specification of requested optimization. The remainder of the input includes multiple image files and a cluster members file. The available options and commands are extensive, and testing the software with the files available for download is an excellent way to gain familiarity with the complex input required. 

Lenstool is referenced in at least 11 manuscripts  \citep [e.g.][]{Julio:2007,Kneib1996,Richard2008,Limousin2007,Richard2009,Jullo2009}. These manuscripts are listed on the Lenstool website \citep{LenstoolWeb}, and many of them include downloadable files of the lens models used in the research. Strong lens galaxy clusters are modeled with parametric methods, and ranked using Bayesian evidence. Although Lenstool was initially developed in 1993 with a downhill $\chi^2$ minimization, modeling of complex systems become inefficient due to the sensitivity of the technique to local minima. The computational method used by Lenstool was then changed to use a publicly available Markov Chain Monte-Carlo sampler, avoiding local minima in the likelihood functions. The merits of this method on simulated strong lensing clusters is demonstrated by Jullo et al \cite{Julio:2007}. 

Using a multi-scale model with a hybrid approach of LTM and non-LTM modeling in the Lenstool software, Jullo and Kneib were able to model Abell 1689, but only for a limited subset of images \citep{Jullo2009}. The key feature is that Lenstool uses a multi-scale model, allowing sharper contrast in regions of higher density. This arrangement of potentials of different sizes allows Lenstool to produce a high-resolution model with a minimum number of parameters. In this study, a mass reconstruction was created using Lenstool.  The model combined a grid of radial basis functions and galaxy scale clumps with cluster member galaxies. A grid was built from a mass map based on 2 cluster-scale and 60 galaxy-scale clumps of mass, instead of the 190 galaxy-scale clumps used in previous studies \citep{Limousin2007}. A catalog of 28 images in 12 systems of multiple images were selected for this analysis. There were 122 parameters, which took 15 days to produce 2000 MCMC samples on a 2.4GHz processor. The results of this study confirmed the ability of a multi-scale model to be used as a lens model. The authors report that the errors between the positions of observed and predicted images were halved, compared to previous studies.

The nature of the mass distribution in Abell 1703 was studied using Lenstool \citep{Richard2009}. This demonstrated the ability to model the inner mass distribution of massive galaxy clusters. This study used a spectroscopic survey to confirm photometric redshifts and precisely constrain the mass distribution in Abell 1703.  Lenstool was used to constrain a parametric mass model with the identified multiple systems. The positions of the multiply imaged systems were used to optimize parameters describing the mass distribution, using a Pseudo Isothermal Elliptical Mass Distribution (PIEMD) using the profile derived from photometry. The Bayesian approach in Lenstool provides a large number of models which sample the probability density function of all the parameters. Results were compared to previous weak-lensing studies, and were found to have a very good fit. This strong lensing analysis using Lenstool and a simple NFW component for the large scale dark matter distribution, was able to reproduce the large number of images in Abell 1703, as well as demonstrate consistency with previous weak-lensing analyses.

\subsubsection{Summary}
Lenstool is a comprehensive program for  gravitational lens modeling. The software is relatively easy to install and use, and is accompanied by extensive documentation. It has been extensively used in the modeling of  observed gravitational lenses. The availability of the input files from previous studies  increases the utility of this software. The software uses a novel technique to combine the strengths of both LTM and non-LTM models, by using a multi-scale model that allows sharper contrast in areas of high density. The flexibility of this approach allows improved prediction of image positions compared to previous studies.

\subsection{LensPerfect}

\subsubsection{Software features}

\begin{figure}[tp]
\includegraphics[trim = 10mm 10mm 10mm 5mm, clip, width=\linewidth]{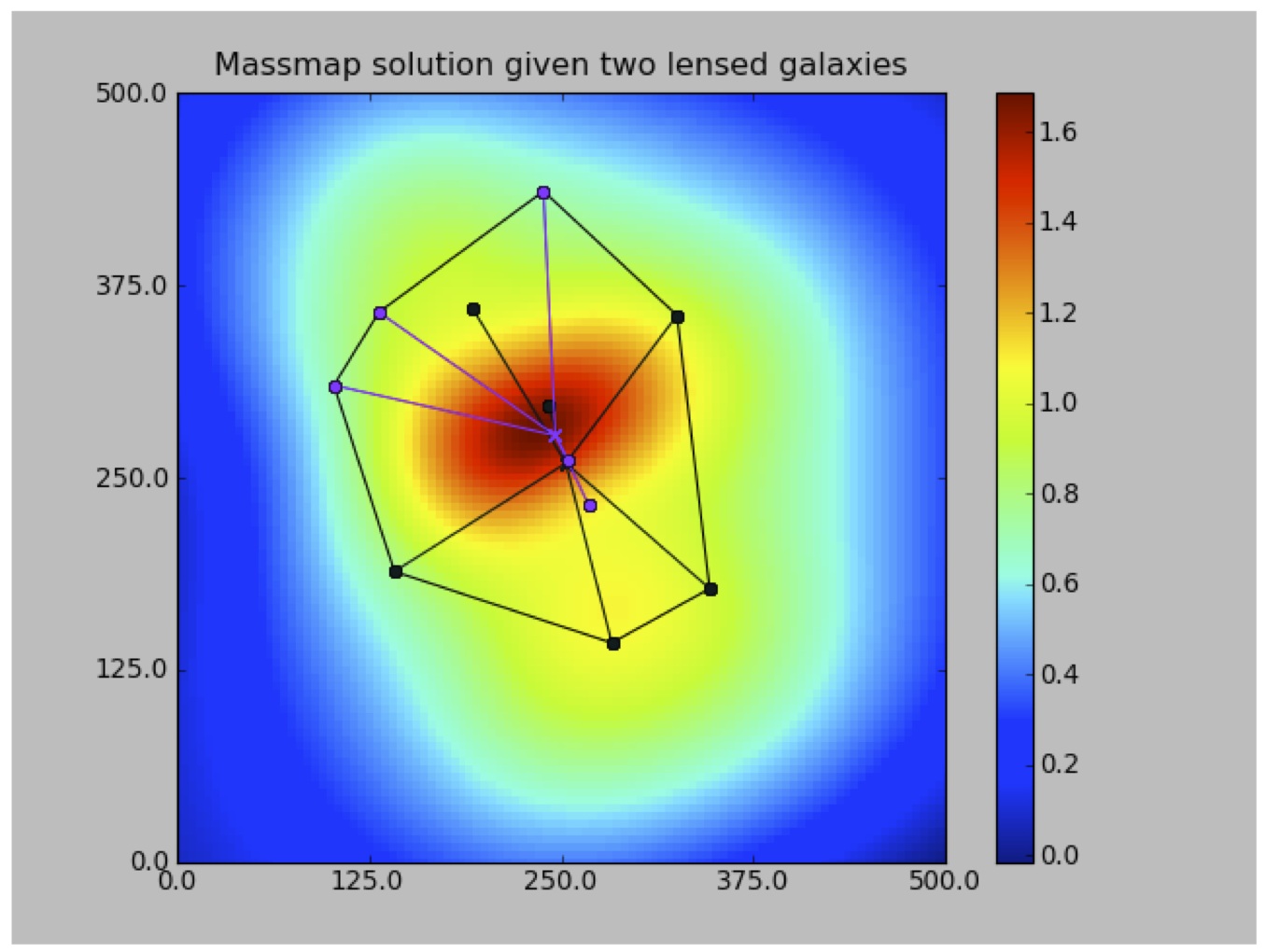}
\caption{Sample output from LensPerfect using a supplied test dataset}
\label{lpfig1}
\end{figure}

LensPerfect was released in 2007 and is available for download \footnote{\url{http://www.its.caltech.edu/~coe/LensPerfect/download/}}. The software is written in Python and the source code is included. Dependencies include Numpy, Scipy, Matplotlib and Pyfits. Installation of the software is slightly complicated by the fact that the software will not work with the latest version of Python.  There are numerous sample data files provided to demonstrate the features of the software. Output from analysis of one of the sample files is shown in Fig \ref{lpfig1}. Documentation is available on two web sites including one which details installation and use of the software \citep{LensPerfectWeb}, although there is no separate user manual. The user interface is character based and input is through a text file with text commands.  

\subsubsection{Lensing features}
The software was developed by Coe  \citep{Coe2008}  and has also been used in one more research study \citep{Coe2010}. LensPerfect uses a parametric model but is also ``model-free'' as described by its developers, who further characterize it as non-LTM. LensPerfect solutions are given as sums of basis functions. While most parametric models are based on a physical object, the basis functions used by LensPerfect have no physical interpretation. Input to the program is via a text file and graphical output is shown immediately on the display. 

LensPerfect represents a new approach to gravitational lens mass map reconstruction, and is the first method to do so using strong gravitational lensing data (multiple images). This new approach uses direct inversion to obtain assumption-free mass map solutions which perfectly reproduce all multiple image positions. This was developed using a new mathematical approach, using a curl-free interpolation of vectors given at scattered data points. One of the key features in any model is the measure of "physicality" of the model. The developers of LensPerfect use a new measure of physicality, with the following traits: (i) positive mass everywhere within the convex hull, (ii) low mass scatter in each radial bin, (iii) no ``tunnels'', (iv) overall smoothness and (v) average mass in radial bins decreases outwards. The only rigid constraint among these is the first trait, requiring positive mass. A complete discussion of these traits is available in the reference \citep{Coe2008}.  

LensPerfect provides an accurate mass map even when there are many lensed galaxies, by using several novel approaches. A weighted average of predicted source positions is used to determine each new source position. The solution is rebuilt at each iteration as new sources are added. This process is fast, and results in an accurate mass map.  Both source and image positions are always perfectly constrained. 

The galaxy cluster Abell 1689 is one of the most studied gravitational lens systems, and thus is ideal for comparisons among lensing models. The large number of multiple images in Abell 1689 also make detailed analysis a challenging problem.  In a followup investigation after introducing the method and software, LensPerfect was applied to the analysis of Abell 1689 using the positions of 168 multiple images. The non-LTM models from LensPerfect were able to reproduce the observed input positions of 168 multiple images of 55 knots residing within 135 images of 42 galaxies \citep{Coe2010}. The computing problem associated with analyzing Abell 1689 is obvious, since it has 100+ strong lensing features. The software must produce a mass model with correct amounts of mass to deflect light from 30+ background galaxies into multiple paths to arrive at the 100+ observed positions. LensPerfect did this, using direct matrix inversion to find solutions based on the input data. The optimization took two weeks to run on a Macbook Pro (Apple Corp., Cupertino CA) laptop computer. This study demonstrates the robustness of the algorithm and its computing efficiency. 

A mass map of Abell 1689 produced by LensPerfect using NFW and S\'{e}rsic fits had a recovered mass profile which matched the input mass profile extremely well \citep{Coe2010}. The NFW fit parameters compare very well to previous studies of Abell 1689 using strong lensing data as well as studies using a combination of strong and weak lensing data. 

\subsubsection{Summary}
LensPerfect is a full-featured program that is extremely capable of producing a mass map based on lensed images, even when there are many galaxies as demonstrated in the followup study of Abell 1689. It is distributed as source code, written in Python, which is available for many platforms. The new approach taken by LensPerfect results in fast and accurate modeling, even of complex systems. The output is graphical and displayed directly. LensPerfect does an excellent job of accurate reconstruction of the sources and lensed images. The ability of LensPerfect to accurately produce a mass map using strong lensing data represents an important advance in gravitational lens research, and will likely lead to further advances.

\subsection{glafic}

\subsubsection{Software features}

The current release of glafic was written in 2012, and is provided as a downloadable executable file which is unpacked from a .tar file \footnote{\url{http://www.slac.stanford.edu/~oguri/glafic/}}. Several sample input files (also in a single .tar file) and a detailed 51 page Users manual are provided. Source code is not provided. There is no installation procedure since it is provided as an executable file for OS X. The program runs without any modifications necessary. The user interface is character based and input is through a text file with text commands. The program runs in the command line interface using OS X. There is no graphical interface. 

\subsubsection{Lensing features}
\begin{figure}[tp]
\includegraphics[width=\linewidth]{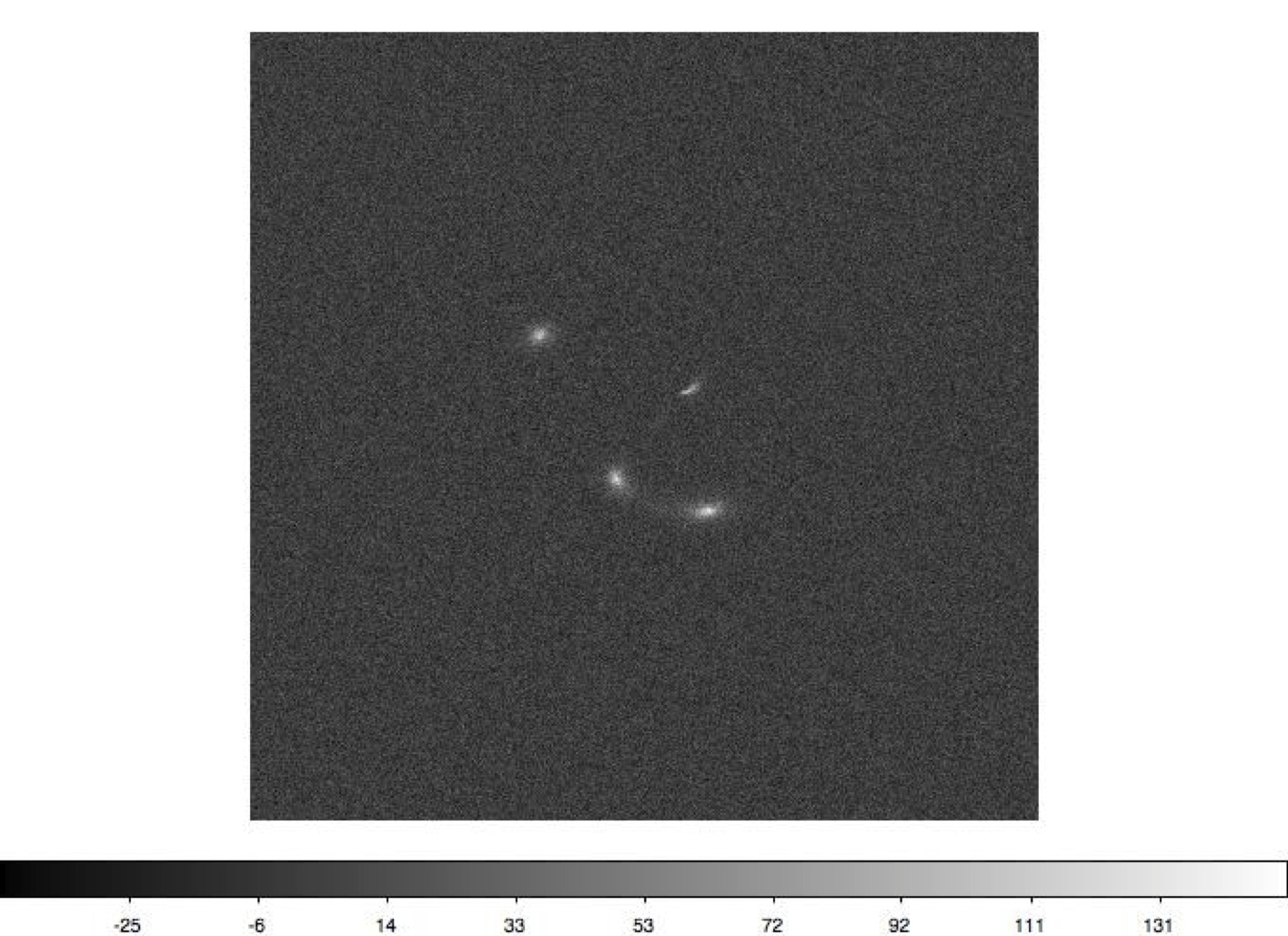}
\caption{FITS image output by glafic for a sample model with a single point source, two lenses and two extended sources}
\label{GlaficFig}
\end{figure}

Glafic uses a parametric model that can be used for a wide variety of gravitational lensing analyses \citep{Oguri2010}. It includes computation of lensed images for both point and extended sources, handling of multiple sources, a wide variety of lens potentials and a technique for mass modeling. Commands are entered in a simple text file, which begins with a list of primary parameters (omega, lambda, Hubble, lens redshift z, pixel size, etc.) and then an optional list of secondary parameters (optional data files, output format desired, extended source normalizations, etc). Point sources are defined simply by their redshift and x- and y-coordinates. Each lens is defined by the lens model and seven parameters. A large catalog of lens models is available (including point mass, Hernquist, NFW, Einsato, S\'{e}rsic, etc.). Extended sources can be Gaussian, S\'{e}rsic, top hat, Moffat or multiple sources. After defining the parameters and the lens models, parameters to be varied in the  $\chi^2$ minimizations are specified. Following this, the desired commands are issued such as computing various lensing properties, Einstein radius, write lensing properties to a FITS file (see sample output in Fig \ref{GlaficFig}), etc. Various types of optimization are permitted. The commands are well described and illustrated in the User's manual. Commands can be entered as a batch using an input file, or entered on a command line. A number of sample data files are provided which illustrate a number of the major features of glafic.

Glafic was used to perform a strong lens analysis of SDSS J1004+4112 \citep{Oguri2010}. This is a particularly interesting quasar lens because it is one of only two known examples of a cluster-scale quasar lens, and contains multiply imaged galaxies at z $\sim$  3. The authors include an indirect comparison with multiple previous mass models of this interesting lens. This study used a parametric model, with the main halo of the lensing cluster modeled with the generalized NFW profile. A standard $\chi^2$ minimization was used to find the best-fit mass model. The best-fit radial mass profile generated is in good agreement with strong lensing data inferred from Chandra X-ray observations. The model used several new constraints including positions of spectroscopically confirmed multiple imaged galaxies, time delays between quasar images, and faint central images. The model thus generated was able to successfully reproduce all observations including time delays.

\subsubsection{Summary}
The glafic program is a well-designed, flexible and easy to use lens modeling program. It includes a wide variety of lens models and is extremely flexible in the types of models it accepts. The User's manual is helpful and complete. Output consists of a variety of possible data files including FITS images. Glafic can accurately recover lens model parameters of known lensing systems.

\subsection {PixeLens}

\subsubsection{Software features}
PixeLens was written in 2007, and is available to run in a web browser or as a stand alone java applet which can be downloaded from the website \footnote{\url{http://www.qgd.uzh.ch/projects/pixelens/}}. There is no installation needed. This program was used at the ANGLES School for Gravitational Lensing in 2007. The source code is  provided in the downloadable .jar file.  Information is also available on the website. Sample input data is provided. A paper (19 pages) and tutorial (14 pages) detailing the software and underlying theory are available on the website, and serve as excellent documentation with numerous examples. The user interface is a single window with several entry panels in the window.  Input is done in the window, or through an input file. 

\subsubsection{Lensing Features}
\begin{figure}[tp]
\includegraphics[width=\linewidth]{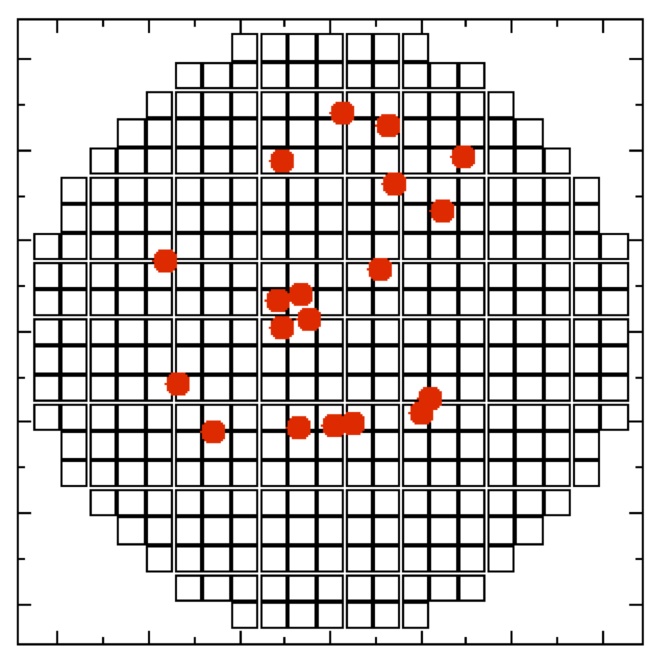}
\caption{Sample output from PixeLens}
\label{PixelensFig}
\end{figure}

This program uses a non-parametric model, with Bayesian statistics. Input to the program consists of model constants (red shifts, pixel size, etc) and image data. The radius of the mass map in pixels and the redshifts for the lens and source must be given.  Optional inputs include setting the mass map as symmetric or asymmetric, the radius of the mass map, external shear, the number of models allowed, and several others. Image data includes x- and y- locations and the time delays. Type of output can be selected as text or eps files. Sample output is shown in Fig. \ref{PixelensFig}, showing the image directly on the display. 

\cite{Saha2006b} used PixeLens using non-LTM models to perfectly reproduce some of the data for Abell1689, but computational limitations restricted PixeLens to fitting 30 multiple images at a time, because of the requirement for exact fits to the image data. This could be considered a virtue allowing the data to be split into two sets as pointed out in the study. In this study, PixeLens had been enhanced to use multiple-source redshifts. PixeLens  had some advantages in the models generated, particularly in handling the inherent problem of degeneracy. While some software uses one or a few models, PixeLens generates a large ensemble of models which explore the possible mass distributions that can reproduce the data. Data from PixeLens was compared to independent datasets for consistency \citep{Saha2006b}. PixeLens has also been used to model the giant quadruple quasar SDSS J1004+4112 \citep{Saha2004,Saha2006b}. In addition to the lensing data (image positions), six kinds of constraints were applied to limit the ensemble to lenses that could plausibly be galaxies or clusters. The result was free-form reconstructions allowed detection of structure in the lens associated with cluster galaxies \citep{Saha2004}. J1004+4112 was reconstructed using 13 images from 4 sources. 

\subsubsection{Summary}
PixeLens is an easy to use program for non-parametric modeling of gravitational lenses. There is ample documentation available. It is based on reconstruction of a pixellated mass map by generating large ensembles of models with a  Metropolis algorithm. It can model several lenses simultaneously, which is a rare feature of modeling software. The code has been tested with a number of fake models and correctly recovers both the mass distributions of the lens and H0 within uncertainties. PixeLens has been used in several published studies to analyze gravitational lenses \citep{Saha2006a,Saha2006b}.

\subsection{SimpLens}

\subsubsection{Software features}
SimpLens was written in 2003, and is available to run in a web browser or as a stand alone java applet which can be downloaded from the website  \footnote{\url{http://www.physik.uzh.ch/~psaha/lens/simplens.php}}.  There is no installation procedure needed. The source code is  provided in the downloadable .jar file. The user interface is a single window which displays graphs of mass, potential and arrival time. There is no formal written documentation, but the web site provides adequate information. 

\subsubsection{Lensing features}
\begin{figure}[tp]
\includegraphics[width=\linewidth]{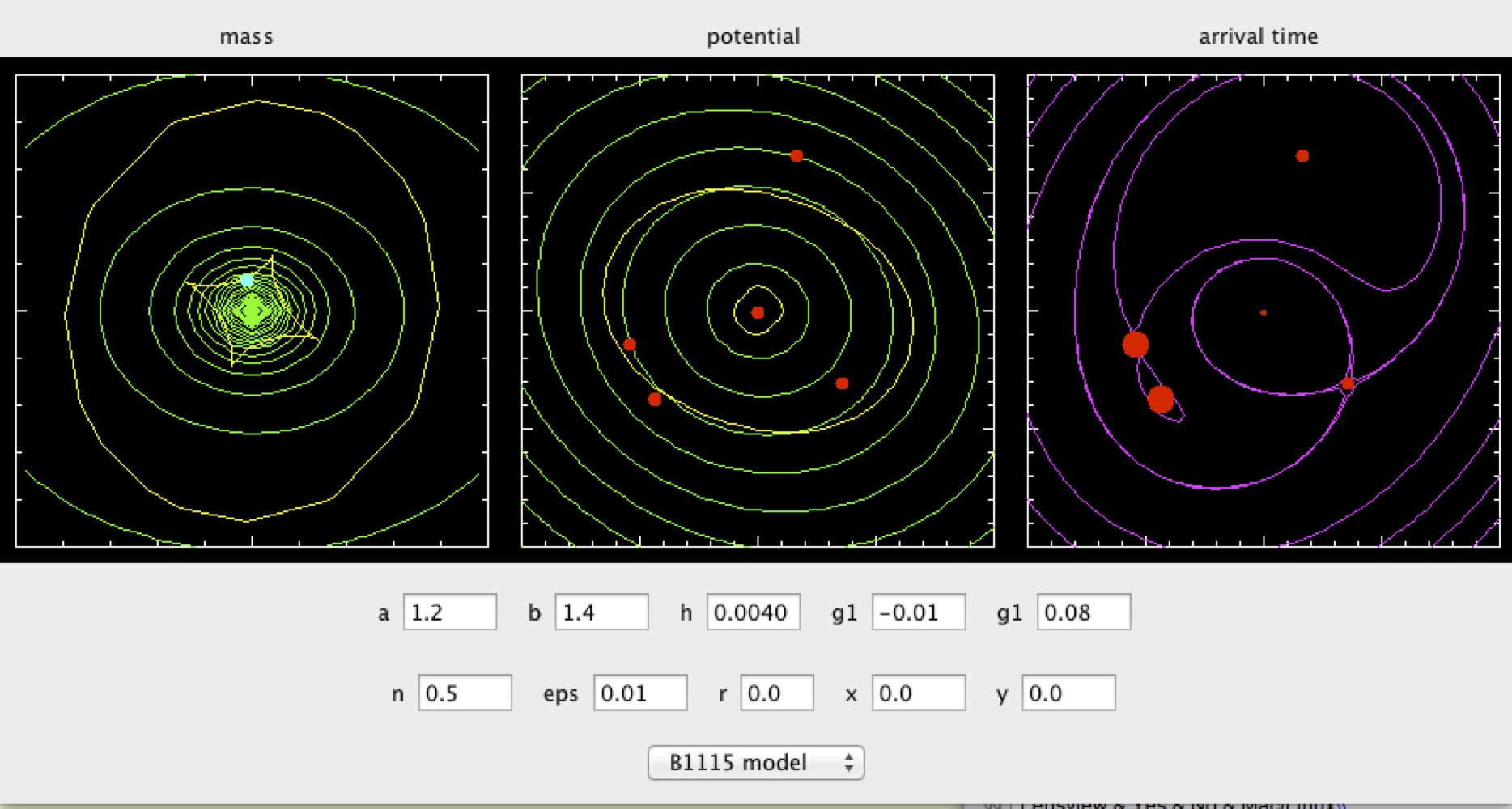}
\caption{Sample data entry and output from SimpLens}
\label{SimplensFig}
\end{figure}

SimpLens uses a  non-parametric model, and the algorithm  is explained in an article written by the  developer \citep{SimplensSaha2003}. That article also demonstrates some of the analytic capabilities of SimpLens. The user can enter values for a,b,h,g1, g2, n, eps, r, x, and y. The parameters a, b, and h specify an elliptical mass distribution, while g1 and g2 refer to the external mass. By clicking the computer mouse, the source position can be changed (see Fig \ref{SimplensFig} left panel) and the effect of that change instantly visualized on the display. The graphs display caustics, saddle point contours and critical curves. A number of sample data sets are provided which are easily selected in a pull-down menu. The various parameters from the sample data can then be varied to instantly observe the effect of the alteration on the three graphs. For example, the sample data allows one to easily observe the effect of added shear.

\subsubsection{Summary}
SimpLens is a simple interactive program which allows one to instantly demonstrate the effect of changing parameters on caustics, saddle points and critical curves for mass, potential and arrival time. It is very easy to use and very instructive.

\subsection{Lensview}

\subsubsection{Software features}

Lensview was first released in 2006 and he current version 1.1.2 is dated 2008  \footnote{\url{http://cira.ivec.org/dokuwiki/doku.php/staff/rwayth/lensview}}. This software is distributed as source code, and has dependencies on CFITSIO, GSL and the FFTW libraries, as well as an optional dependency on fastsell. It can be installed in Linux or OS X. There is a web site where the software can be downloaded, with information regarding installation and use, but there is no separate user manual \citep{Lensview}. The user interface is character based and input is through a text file with text commands. The software is described as running in two modes, "simple projection" and "normal" (fit an observed image).

\subsubsection{Lensing features}
Lensview is based on the LensMEM algorithm \citep{LensMEMAlg}, which finds the best fitting lens mass model and source brightness distribution using a maximum entropy constraint \citep{Wayth2006}. Lensview is used to study lens inversion, obtaining a model of the source based on lensed images. It uses a parametric model for solving the lens inversion problem. The main features of the software include: (i) projection between image and source plane conserving surface brightness, (ii) compound lens models which can contain several basic components, (iii) reconstruction of the unlensed source brightness profile and corresponding model image using a non-parametric source, (iv) iterative source reconstruction process incorporating a maximum entropy metric, and (v) statistical evaluation of the model image given by the current lens model parameters. 

In simple mode, Lensview takes as input: a source, a lens model, a data image (optional) and projects the source onto the image. The projected image and $\chi^2$ are output. A log file is also produced. A variety of lens model components are supported, including pseudo-isothermal potential (PIEP), SIE, point mass, power-law mass distribution, NFW, constant density mass sheet, external shear, exponential disc and a S\'{e}rsic mass distribution. Sample templates for each of these mass distributions are given on the website \citep{Lensview}.  To use the software, one needs a data image, a PSF image and a mask image, all of which need to be the same size. There are no sample data files provided with the software, but there are some examples available through the website. Output from the program can be a FITS image, or a file for input into IDL and plotting the $\chi^2$ surfaces. 

In their study, Wayth and Webster first used Lensview with simulated optical images to determine optimal pixel size, accuracy of lens model parameters, distinguishing power of the image and overall goodness-of-fit \citep{Wayth2006}. Model images were created with a two-component source having two regions of different size and peak brightness. A PSF was generated and convolved with the simulated image. They found that an image to source pixel scale ratio of at least 2 is required to reproduce the data and a ratio of 3 was optimal. A sufficiently large source plane of approximately 15x15 pixels was adequate. All lens model parameters were well recovered using this data. A comparison of various models showed that the image could distinguish between lens models although the differences were quite small.

Lensview was used to analyze the Einstein Ring ER 0047-2808 \citep{Wayth2005}. This study modeled the system using six different models, including PIEP, SIS+$\gamma$, SIE, SPEMD, M/L and NFW. This study showed that the SIS+$\gamma$, M/L and NFW (25kpc) could not reproduce the data, but that the other models could. The differences between the other, successful models, were subtle. The data showed that the lensed image has four distinct bright regions, and that Lensview was able to generate a best-fitting image and reconstructed the source brightness profile using a non-parametric model. 

Lensview was also used to analyze the optically lensed arc HST J15433+5352 \citep{Wayth2006}. This lens was modeled with a PIEP model, including an external shear. The final source plane used was 10x10 pixels, with a source-to-image plane pixel scale ratio of 1/2. The resulting model was somewhat surprising in that a purely elliptical model reproduced the data equally to a model including shear. They found a critical radius (b) of 0.525, which differed somewhat from data previously reported with a critical radius of 0.58 \citep{Knudson2001}. Some of this difference could be accounted for the fact that the Lensview study is a non-parametric model, while Knudson et al \cite{Knudson2001} used a parametric model. 

Lin et al used Lensview to study a bright strongly lensed galaxy in the SDSS DR5 \citep{Lin2009}. Lensview provided an excellent model, but as the authors pointed out, Lensview uses the full image information which precludes determination of how well the image positions are determined. They then used Lensmodel to fit an SIE model using only the image positions. This study illustrates both the strengths and weaknesses of this approach, but more importantly also suggests the need for using multiple approaches to modeling.

\subsubsection{Summary}
Lensview is a comprehensive modeling program, with a large number of features and options. It is not straightforward to use, although the sample data files and suggestions provided on the website do facilitate gaining proficiency with the software. Due to its comprehensive nature, it is possible to specify very complicated lens models based on one or more components.

\subsection{GRALE}

\subsubsection{Software features}

The current version of GRALE is 0.9.0, released in 2008 \footnote{\url{http://research.edm.uhasselt.be/~jori/page/index.php?n=Physics.Grale}}. It is provided as source code which must be compiled and linked using the CMAKE utility. The installation procedure is fairly straightforward, and the GSL and CFITSIO libraries are necessary. The software can be installed in Linux, or OS X. The GRALE library is easily run under GRALESHELL, which provides an interactive environment. The software runs in a single window and text commands are entered in a panel at the bottom of the window. Output is  obtained using GNUPLOT (see Fig. \ref{GRALEFig}) or optionally as a FITS image. Documentation consists of a website dedicated to this software, but there is no separate user manual. 

\subsubsection{Lensing features}
\begin{figure}[tp]
\includegraphics[width=\linewidth]{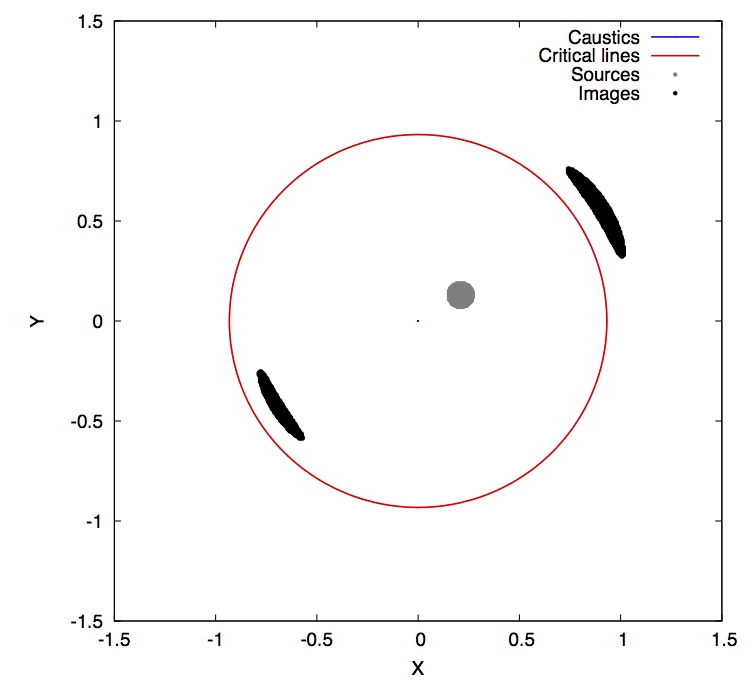}
\caption{Output from GRALE for one of the supplied sample data files}
\label{GRALEFig}
\end{figure}

GRALE can be used to simulate gravitational lenses and to invert lensing systems. The GRALE algorithm uses a non-parametric technique to infer the mass distribution of a gravitational lens system with multiple strong lensed systems \citep{Liesenborgs2006}. To start simulating a gravitational lens, the user first decides which type of lens to use. There are a variety to select from including a point mass lens, a SIS, SIE, projected Plummer sphere, square shaped region of constant density, a two dimensional gaussian density profile, and others. The lens parameters include mass, distance and others as appropriate. Commands and parameters are entered line by line in GRALESHELL. Next, the distances to the source plane are specified and then the mapping from the image plane to the source plane is specified. Sources can then be added. A sample lens model is shown in Fig. \ref{GRALEFig}. Lens parameters can be optionally saved and retrieved. The program directly outputs a file which is used by GNUPLOT to generate visual output. 

GRALE  has been used in a number of published studies to analyze various lenses.  GRALE functions well in modeling systems with few images, and thus less information  \citep{GRALE2}.  GRALE was used in a study using strong lens modeling to search for dark matter \citep{GRALE4}.  

GRALE was used to model the system Cl 0024+1654, and infer the mass maps using a non-parametric technique \citep{GRALE3}. This represents the first time that image information alone was used to reconstruct the mass distribution of this cluster in the strong lensing region. No information about the positions of cluster members was used. Image data for sources A and B as previously described was used for the inversion procedure to reconstruct the source. Source A was mapped onto five images and source B mapped onto two images. A mass map was obtained by averaging 28 solutions. The only bias in the technique is that the user must specify a square shaped area for the algorithm to search mass distribution, assuming that no mass is outside the boundaries of that region. Using the inversion procedure in GRALE, an averaged mass map was obtained that displayed the features seen in the ACS images. 

Models of the well known system SDSS J1044+4112 were also created using GRALE \citep{GRALE5}. This study looked at five spectroscopically confirmed images and position information was used from existing studies. One of the images was uncertain (A5) and was not included in the first inversion. A second inversion was then performed with the addition of image A5, which allowed reconstruction of the source shapes after projecting back onto the source planes. Calculation of the total mass within 60Kpc and 110Kpc compared well to results in existing studies. Comparison was also made to the best fit NFW model, and the configuration of the cluster corresponded to that reported previously \citep{Oguri2004}.

\subsubsection{Summary}
GRALE is easily installed and straightforward to use. Commands are text-based and relatively intuitive. A variety of lens models is available for simulations. A large number of lens and source parameters can be entered by the user, allowing one to perform complex lens simulations. In several studies, GRALE did an excellent job of recovering lens parameters and accurately reconstructing the source parameters. GRALE has been tested with models of C 0024+1654 and SDSS J1004+4112 and gives results compatible with previous studies.


\section{Educational Software}\label{Educ}

Programs suited for education are useful to demonstrate the basic principles of lensing. Categorization as educational software is  arbitrary and based on the fact that there were no studies found in the literature that used these four software packages for data analysis in published studies.

\subsection{GravLens HD}

\subsubsection{Software features}
GravLensHD is available for iOS devices and is obtained  from the Apple (Cupertino CA) App Store, at no cost \footnote{Available on iOS devices though the App Store (Apple Co. Cupertino CA) application}. The source code is not provided.  There is a help-screen with some suggested exercises to demonstrate the features of a gravitational lens. The web site has some additional information \citep{GravLensHDWebsite}. There is no formal documentation. The interface is limited to a single touch screen. Installation is  simple using the App Store application. 

\subsubsection{Lensing Features}
This software uses a single mass as a lens, and is categorized as a parametric model. The software uses a simulated source and a foreground lensing mass to draw arcs or an Einstein ring as one moves the lensing mass relative to the star digitally on the touch screen. The lensing mass can be made invisible leaving the lensed image of the source. The ellipticity of the lensing mass can be changed through a few preset shapes. The size of the lensing mass can be changed with on-screen gestures. The images can be saved to the devices library.The background image can be changed to any picture of the user's choice, giving an excellent way to show the effect of a lens on a known image. Some technical material is also available within the application. It uses a singular isothermal sphere (SIS) model with external shear. The maximum shear permitted is 0.6.

\subsubsection{Summary}
In summary, this software provides an excellent demonstration of the effect of a gravitational lens, and may serve as a tool to stimulate young minds to think about astronomy. The concepts used are the same as those for any gravitational lensing system. There is no option for  numerical input, and all interaction with the software is through the touch screen. Further information is available on a web site dedicated to this software \citep{GravLensHDWebsite}.

\subsection{G-Lens}

\subsubsection{Software features}
G-Lens  was written by Boughen in 1998 (as part of an undergraduate honors thesis), and is available as an on-line download as a single executable file  \footnote{\url{http://astrotips.com/software/g-lens}}. The program will only run in a DOS or early Windows environment, but can be used with more modern operating systems in DOS emulators, such as DOSBox which runs in OS X \citep{Dosbox}. Source code is not provided. There is no installation, as one needs only to execute the .exe file provided. The software was reviewed in the June 2000 issue of Sky and Telescope \citep{Glens}. Along with the software, there is a one-page instruction sheet available, but no other documentation. The user interface is a single character based text box where the values of parameters are entered. 

\subsubsection{Lensing Features}
The lens mass is varied and is modeled as a single point mass, characterizing this software as a parametric model. Input to the program is performed by providing the lensing mass (in solar masses), the lens and source distances (in Mpc)  and then selecting among three options for geometry of the lens (circle, ellipse and grid). Output is immediately shown on the  display. The image can be optionally printed. 

\subsubsection{Summary}
This software simulates a simple point mass lens and allows numerical input regarding basic  system geometry. The output is graphical showing the image resulting  from the lens designed by the user. The ability to specify basic parameters numerically makes this a good program for demonstration of basic lens effects.

\subsection{Gravitational Lensing}

\subsubsection{Software features}
Gravitational Lensing was written in 2002 and is distributed as an executable file which runs in Windows, and also as a C program for use on an HP-49 calculator. The software is available for download on the website \footnote{\url{http://www.kwakkelflap.com/gravlens.html}}.  There is no installation, as one needs only to run the .exe file provided. Source code for the Windows version is not distributed. Documentation is available to explain the science of lensing, written in Dutch. There is no documentation for running the software. The interface is a single window, and a ``File'' menu. The program has an ``About'' screen but no on-line help. The options in the ``File'' menu allows changing parameters of the lens model. After changing the parameters, the result is displayed in the window.

\subsubsection{Lensing features}
\begin{figure}[tp]
\includegraphics[width=\linewidth]{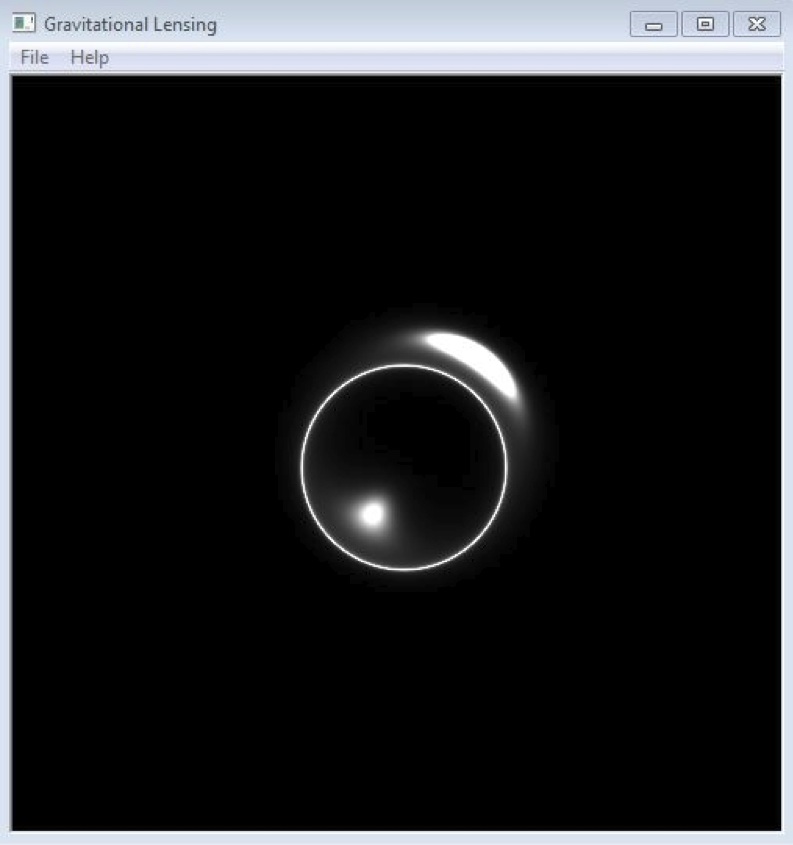}
\caption{Sample output from Gravitational Lensing}
\label{WinGravlens}
\end{figure}

The mass can be varied on the data entry screen and is modeled as a point mass, characterizing this as a parametric model. Sample output is shown in Fig \ref{WinGravlens}. The mass of the source and lens can be set individually (as multiples of solar masses). Distances from the observer to source and lens are set in Mpc. The position of the source can be set as an offset from center, and the angle of view can be varied. Models include a Plummer model or isothermal sphere. The Einstein radius can be optionally set.  A set of default values is also available to show the capabilities of the program.

\subsubsection{Summary}
This simulation is simple to install and use. Various parameters are easily set in a menu screen and the resulting lens model instantly visualized. There are two optional models that can be used. This software is very useful as an educational tool, but not sufficiently robust for research analysis.

\subsection{lens}

\subsubsection{Software features}
Lens was released in 2002, and is written in MATLAB, and therefore the MATLAB (The MathWorks Inc, Natick MA USA) software is required. The source code is  distributed and easily installed in MATLAB (these tests conducted with version R2010a). The source code and sample data files (SIS1 and SIS2) are available for download \footnote{\url{http://www.astro.ubc.ca/people/newbury/siam/lens.html}}. The package includes two programs written in MATLAB, lens (``forward'' modeling) and invert (which solves the lens inversion problem). The information on the website includes a tutorial, and serves as excellent documentation. The user interface is through the MATLAB screen. Installation is accomplished using MATLAB. Sample data files are provided to test the software.

\subsubsection{Lensing features}
\begin{figure}[tp]
\includegraphics[width=\linewidth]{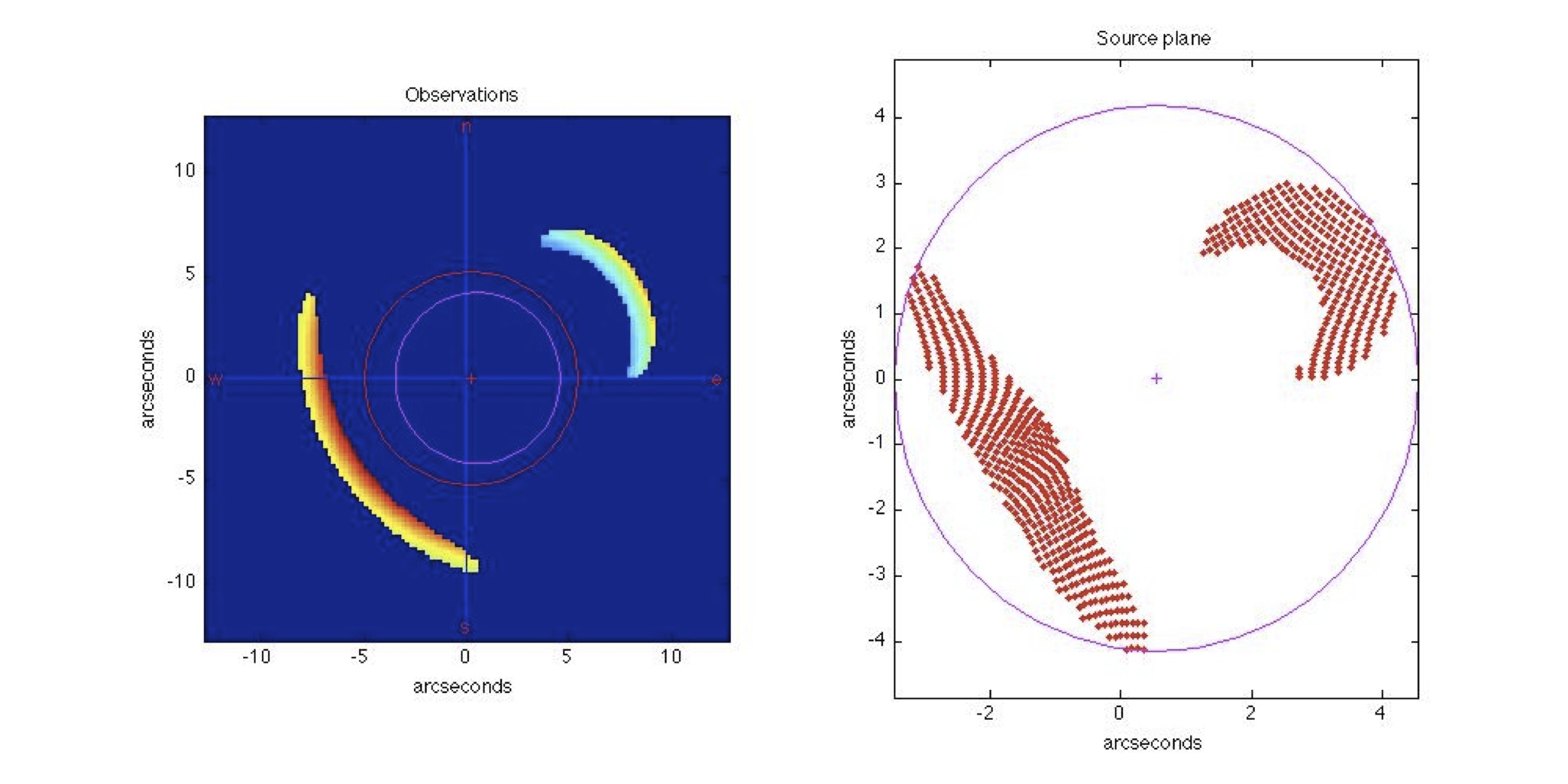}
\caption{Sample output from lens showing the image plane (left) and the generated source plane (right)}
\label{SIS2}
\end{figure}

Lens is a hybrid program, which demonstrates both the ``forward'' problem as well as lens inversion. For modeling the lensed images, lens uses a parametric model as a SIS, and the algorithm is described in an accompanying article \citep{LensArticle} which is easily available following a link from the website. Various model parameters of the SIS can be changed in the parameter file, as detailed on the website. The input data file specifies observation parameters, lens parameters (sigma, z) and source parameters (z,x,y, ellipticity, flux) and produces the lensed image (the ``forward'' problem") (left panel, Fig \ref{SIS2}). Resulting images are shown directly on the display with no further intervention. The program optionally writes a data file which can then be used as input to the inversion routine to solve the ``reverse'' problem. The user estimates the center of the lensing mass with the computer mouse, and an image of the source plane is created (right panel, Fig \ref{SIS2}).

\subsubsection{Summary}
This program is easy to use, but requires MATLAB. The program demonstrates a solution to the basic problems of gravitational lensing including the forward and reverse problems. Input files are easy to create and the output graphics are quickly generated and illustrate the solution. The software could be easily modified, and is supplied with ample comments.

\subsection{MOWGLI}

\subsubsection{Software features}
MOWGLI is available as a Java applet, accessed through a web browser \footnote{\url{http://www.ephysics.org/mowgli/}}. The web site includes a Quick-start guide on-screen with easy to follow directions. The interface is intuitive and requires minimal training to use. There is no documentation as all information is described on the web site. 

\subsubsection{Lensing features}
MOWGLI is designed for interactive manual modeling of strong gravitational lenses, as described on the web site. Objects such as lenses, sources or masks are added to the screen through an easy to use interface. One can upload a .jpg image of the system for which modeling is desired. By clicking and dragging on the screen, the objects can be moved into various positions as desired. Once the objects are positioned, there are two buttons labeled 'predicted' and 'residual' to view these two images on the right hand panel of the interface. The residual can be minimized by adjusting the Chi square with an on-screen button. 

\subsubsection{Summary}
This software is easy to use and easily accessible as a Java applet. According to the website, a paper has been submitted to further describe the features.

\begin{table}[t]

\centering
\begin{tabular}{c c c}
\hline
Package & Model & Algorithm \\
\hline
gravlens / lensmodel & Para & Image tiling \\
Lenstool & Para/Non-Para & MCMC \\
LensPerfect & Non-Para & Vector interpolation  \\
glafic & Para & Adaptive meshing \\
PixeLens & Non-Para & Pixelated mass map \\
SimpLens & Non-Para & - \\
Lensview & Non-para & LensMEM \\
GRALE & Non-Para & Genetic  \\
\hline
GravLensHD & Para & - \\
G-Lens & Para & - \\
Gravitational Lensing & Para & - \\
lens & Para/Non-para & - \\
MOWGLI & - & - \\

\end{tabular}

\caption{Lensing features of gravitational lens modeling software. The type of lens models and algorithm used by each of the software packages is shown. Para=parametric (LTM), Non-Para= Non-parametric (Non-LTM)}

\label{table:Lensing}
\end{table}

\section{Discussion} \label{Compare}
Most research using gravitational lens modeling software utilizes a single software package to model an observed lensing system. Results are then compared with observational data or other models from the literature. This is the first review to specifically examine strong gravitational lens modeling software. This information is of particular importance because there are no single standard programs used for modeling gravitational lenses.

\subsection{Software Selection}
There are a number of factors which are important when selecting appropriate software. Investigators wishing to develop their own may use existing software, especially if source code is provided, as a starting point for feature sets, data entry, etc. The computing platform may also be an important factor, although most of the research software reviewed here will run on a Linux system. Some of the software reviewed runs on a limited number of systems. The type of data used for input and the results provided are also important factors. Some of the programs provide immediate display of output while others require another step, with other software, to display the results. 

\subsection{Software Comparisons}

 Comparisons of gravitational lens modeling software will be defined as \emph{indirect comparisons} or \emph{direct comparisons}. An indirect comparison is comparing results from various software, in different papers, modeling the same lensing system. This is relatively easy because the software is commonly tested using well-described lensing systems such as Cl0024+1654 (GRALE:\citep{GRALE3}), SDSS J1004+4112 (glafic:\citep{Oguri2010}, GRALE:\citep{GRALE5} and PixeLens:\citep{Saha2004}), and Abell1689 (Lenstool:\citep{Jullo2009}, LensPerfect:\citep{Coe2010}, and PixeLens:\citep{Saha2006b}). Direct comparisons  compare the models generated for a single lensing system using different software in the same publication. However, there are very few direct comparisons of software in the literature.

The modeling of Abell 1689 by PixeLens, LensPerfect and Lenstool has been compared qualitatively \citep{Coe2010}. Lin and coworkers use Lensview to model a bright z=2 galaxy and then compared the results to another model made using Lenstool \citep{Lin2009}.  There is often no single ideal modeling software, as was illustrated in the study by Lin and colleagues who found that the Lensview model made it difficult to determine how well image positions alone are determined \citep{Lin2009}. Further studies are needed to compare the ability of various software packages to model the same lensing system in order to enable meaningful comparisons. Understanding the underlying assumptions and limitations of the various software will be facilitated in the future by more direct comparison studies. 

Future research in gravitational lens modeling will also require direct comparisons of the results of models using data from both strong and weak gravitational lensing analysis. Such analyses have already been reported for Abell1689 \citep{Limousin2007,Coe2010}. 

The use of direct comparisons will enable better comparisons of software usability, feature sets as well as checking results against known software. Future studies in gravitational lens modeling will demand new approaches and sophistication.

\section{Conclusions} \label{Concl}

There is a wide variety of gravitational lensing modeling software available. Many of the publications using these packages are written by the software developers, suggesting that  the software is developed for their personal use. Software available only as executable files has the advantage of being rapidly usable, as long as the computing platform is available. Software distributed as source code may require significant time to compile and prepare given the vagaries of software libraries.

The comprehensiveness of this study is limited by the ability to identify existing software or studies related to the software reviewed. However, a fairly wide spectrum of software was available for review, and this analysis identifies opportunities for improvement based on existing software. 

Awareness of available software may limit the need to develop proprietary software in the future. The source code is available for some applications which  facilitates the development of custom software. Given the increased activity in gravitational lensing research, sharing of software and algorithms may result in significant time savings. The usability of available software is somewhat limited by the essentially consistent use of character based user interfaces. Future software should be modular in nature and use a graphical user interface for improved functionality.

Until recently, the construction of gravitational mass maps to detail dark matter distribution has depended on data from weak lensing. However, as shown in recent studies using LensPerfect, glafic  and GRALE, accurate mass maps can be constructed using data from strong gravitational lensing data, and this may signal a new era in studies of strong gravitational lensing. New approaches to software development will be necessary to support this shift in research, especially with the advent of far more detailed images from the next generation of telescopes. Future studies should include direct comparisons with other available software in addition to indirect comparisons with previous studies of well-described lensing systems. 

\section*{Acknowledgements}

The assistance of  Mark Boughen (G-Lens), Jori Liesenborgs (GRALE), Eric Jullo (LensTool) and Dan Coe (LensPerfect) is gratefully acknowledged.

\bibliographystyle{elsarticle-harv.bst}

\end{document}